\documentclass[fleqn,usenatbib]{mnras}

\usepackage{newtxtext,newtxmath}

\usepackage[T1]{fontenc}

\DeclareRobustCommand{\VAN}[3]{#2}
\let\VANthebibliography\thebibliography
\def\thebibliography{\DeclareRobustCommand{\VAN}[3]{##3}\VANthebibliography}

\usepackage{scalerel}
\usepackage{tikz}
\usetikzlibrary{svg.path}

\definecolor{orcidlogocol}{HTML}{A6CE39}
\tikzset{
  orcidlogo/.pic={
    \fill[orcidlogocol] svg{M256,128c0,70.7-57.3,128-128,128C57.3,256,0,198.7,0,128C0,57.3,57.3,0,128,0C198.7,0,256,57.3,256,128z};
    \fill[white] svg{M86.3,186.2H70.9V79.1h15.4v48.4V186.2z}
                 svg{M108.9,79.1h41.6c39.6,0,57,28.3,57,53.6c0,27.5-21.5,53.6-56.8,53.6h-41.8V79.1z M124.3,172.4h24.5c34.9,0,42.9-26.5,42.9-39.7c0-21.5-13.7-39.7-43.7-39.7h-23.7V172.4z}
                 svg{M88.7,56.8c0,5.5-4.5,10.1-10.1,10.1c-5.6,0-10.1-4.6-10.1-10.1c0-5.6,4.5-10.1,10.1-10.1C84.2,46.7,88.7,51.3,88.7,56.8z};
  }
}

\newcommand\orcidicon[1]{\href{https://orcid.org/#1}{\mbox{\scalerel*{
\begin{tikzpicture}[yscale=-1,transform shape]
\pic{orcidlogo};
\end{tikzpicture}
}{|}}}}
\usepackage{graphicx}	
\usepackage{amsmath}	

\newcommand\msun{M_{\odot}}

\defcitealias{2022Nashimoto}{N22}

\title[UDGs Around MW Analogs]{Extending Ultra-Diffuse Galaxy Abundances to Milky Way Analogs}

\author[Karunakaran \& Zaritsky]{Ananthan Karunakaran$^{1\,\orcidicon{0000-0001-8855-3635}}$\,\thanks{E-mail: ananthan@iaa.es},  Dennis Zaritsky$^{2\,\orcidicon{0000-0002-5177-727X}}$
\\
$^{1}$Instituto de Astrof\'{i}sica de Andaluc\'{i}a (CSIC), Glorieta de la Astronom\'{i}a, 18008 Granada, Spain\\
$^{2}$Steward Observatory, University of Arizona, 933 North Cherry Avenue, Rm. N204, Tucson, AZ 85721-0065, USA\\
}

\date{Accepted XXX. Received YYY; in original form ZZZ}

\pubyear{2022}

\begin{document}
\label{firstpage}
\pagerange{\pageref{firstpage}--\pageref{lastpage}}
\maketitle

\begin{abstract}
We extend the Ultra-Diffuse Galaxy (UDG) abundance relation, $N_{UDG}-M_{200}$, to lower halo mass hosts $(M_{200}\sim10^{11.6-12.2}\msun)$.\ We select UDG satellites from published catalogs of dwarf satellite galaxies around Milky Way analogs, namely the Exploration of Local Volume Satellites (ELVES) survey, Satellite Around Galactic Analogs (SAGA) survey, and a survey of Milky Way-like systems conducted using the Hyper-Suprime Cam.\ Of the 516 satellites around a total of 75 Milky Way-like hosts, we find 41 satellites around 33 hosts satisfy the UDG criteria.\ The distributions of host halo masses peak around $M_{200}\sim10^{12}\msun$ independent of whether the host has a UDG satellite or not.\ We use literature UDG abundances and those derived here to trace the $N_{UDG}-M_{200}$ relation over three orders of magnitude down to $M_{200}=10^{11.6}\msun$ and find a best-fit linear relation of $N_{UDG} = (38\pm5)\cdot(\frac{M_{200}}{10^{14}})^{0.89\pm0.04}$.\ This sub-linear slope is consistent with earlier studies of UDG abundances as well as abundance relations for brighter dwarf galaxies, excluding UDG formation mechanisms that require high-density environments.\ However, we highlight the need for further homogeneous characterization of UDGs across a wide range of environments to properly understand the $N_{UDG}-M_{200}$ relation.
\end{abstract}




\section{Introduction}
Modern wide-field optical surveys have reinvigorated studies of dwarf and low surface brightness (LSB) galaxies.\ A key re-discovery from such work is the population of extended $(\mathrm{effective\, radii,\,} R_{\mathrm{eff}}\geq1.5\,\mathrm{kpc})$, faint $(\mathrm{central\, surface\, brightness,\,} \mu_{g,0}\geq24\,\mathrm{mag\,arcsec^{-2}})$ LSB galaxies dubbed Ultra-Diffuse Galaxies \citep[UDGs,][]{2015vanDokkum}.\ These UDGs are analogous to extended LSB galaxies discovered in early LSB surveys \citep[e.g.][see also \citealt{conselice2018}]{1976Disney,1995Schwartzenberg,1996Impey,1997Dalcanton,1997Oneil}.\ 

Since their detection in the Coma cluster, UDGs have been detected in large quantities across a variety of environments including clusters \citep[e.g.][]{koda2015,2015Mihos,zaritsky2019,lim2020}, groups \citep[e.g.][]{roman2017,muller2018,2020Somalwar,2021Gannon}, and in the field \citep[e.g.][]{2016MartinezDelgado,leisman2017,prole2019,tanoglidis2021,2022Zaritsky}.\ Several proposed UDG formation mechanisms can be broadly placed into two regimes: i) formation via internal processes (bursty star-formation, \citealt{dicintio2017}; high-spin haloes, \citealt{amorisco2016}; weak star-formation feedback, \citealt{2020Mancerapina}), and ii) formation via external processes (tidal interactions, \citealt{carleton2019},\citealt{sales2020}; mergers, \citealt{wright2021}).\ These processes likely depend on the UDG's environment.\ For example, UDGs in low density (i.e.\ field) environments may have formed predominantly through internal processes, while those in more dense environments (i.e.\ groups and clusters) may have formed through either internal processes prior to their in-fall or formed through external processes after their in-fall.\

By estimating the abundance of UDGs as a function of their environment, i.e.\ halo mass $(M_{200})$, we can gain some insight into which processes are dominant in a given environment.\ For example, UDG formation scenarios that require cluster environments can be ruled out if UDGs are more abundant in lower mass halos (i.e.\ groups) compared to higher mass halos (i.e.\ clusters).\

Multiple studies have focused on comparing the abundance of UDGs, $N_{UDG}$, as a function of their host $M_{200}$.\ For example, \citet[][]{2016vanderBurg} presented an early comparison of UDGs in clusters $(M_{200}\sim10^{14-15}\msun)$ and found a near linear $N_{UDG}-M_{200}$ relation with a slope of $0.93\pm0.16$.\ In their follow-up study with a broader sample of host environments $(M_{200}\sim10^{12-15}\msun)$ from the GAMA survey and KiDS, \citet[][]{2017vanderBurg} find a slightly super-linear slope of $1.11\pm0.07$, spanning across three orders of magnitude.\ Confirming a super-linear relation would imply that UDGs are preferentially found or formed within cluster environments and/or are possible more frequently destroyed in groups.\ Investigations by \citet[][]{roman2017} around Hickson Compact Groups and \citet[][]{2019ManceraPina} in galaxy clusters from the KIWICS survey find sub-linear slopes ($\sim0.85\pm0.05$ and $\sim0.81\pm0.17$, respectively), suggesting that UDGs are preferentially found in lower mass host halos.\ These latter studies are consistent with abundances of brighter dwarf galaxies whose relations are also also sub-linear \citep[][]{2009TrenthamTully,2017vanderBurg}.\ In particular, \citet[][]{2017vanderBurg} use the same GAMA and KiDS data to consistently measure a ``mass-richness'' relation for brighter members ($M_r \lesssim -18.7$) in their groups and clusters finding a best-fit relation of $N_{Bright} = (31\pm3)\cdot[\frac{M_{200}}{10^{14}}]^{0.78\pm0.05}$.\ It should be noted that there are ever-growing samples of UDGs and some in relatively nearby clusters and groups \citep[e.g.][among several others]{2016Yagi,2020Forbes,2022LaMarca,2022Venhola}.\ While some of the groups included in these studies extend to lower halo masses (i.e.\ $M_{200}\sim10^{12-13}\msun$), those hosts are far less numerous relative to their higher mass counterparts in the study.\ We take the next step to determine how the $N_{UDG}-M_{200}$ relation extends toward lower host halo masses.\ 

Several efforts focus on the faint satellite populations of nearby Milky Way analogs \citep[e.g.][]{GehaSAGA,ELVESI,2022Nashimoto}.\ These surveys select their Milky Way analogs based on their luminosities (i.e.\ $M_K$), among other criteria, within the nearby Universe ($D<40$ Mpc).\ Satellites are catalogued using extant or newly obtained observations and confirmed either spectroscopically or via Surface Brightness Fluctuation (SBF) distance estimates.\ Crucially, these satellite samples contain several LSBs, some of which may be UDGs.\ 

We exploit these samples to constrain the abundance of UDGs around Milky Way analogs and examine the $N_{UDG}-M_{200}$ relation across a broad range of halo masses (i.e.\ $M_{200}\sim10^{11.5-15}\msun$).\ This paper is organised as follows.\ In Section \ref{sec:sample}, we describe the satellite surveys around Milky Way-like systems and their general properties.\ We present the results from our analysis of the UDG frequency of these systems combined with the results from previous studies in Section \ref{sec:results}.\ We briefly discuss and summarize our results in Section \ref{sec:summary}.

\section{Data}\label{sec:sample}
Here, we briefly describe the samples we use in this work, the selection process of UDGs in these samples, and our methodology for estimating the host masses.\
\subsection{Survey Descriptions}\label{subsec:surveys}
The first sample we use is the Exploration of Local VolumE Satellites (ELVES) survey \citep[][]{ELVESI}.\ The ELVES survey focuses on 31 nearby $(D<12\mathrm{\,Mpc})$ hosts with a broad luminosity range $(-22.1<M_K<-24.9)$ and they are able to identify and catalogue faint satellites $(M_V<-9.5,\, \mu_{0,V}<26.5 {\mathrm{mag\,arcsec^{-2}}})$, as characterized by their artificial dwarf recovery simulations.\ Their photometric completeness is due to the relative proximity of these systems but also the dedicated photometry extracted for these systems.\ In total, there are 338 satellites that have been catalogued and their associations to their hosts are confirmed via surface brightness fluctuation (SBF) distance estimates.\ We note that \citet[][]{ELVESI} identify an additional 106 satellite candidates that are awaiting distance measurements.\ However, we do not include them in our analysis and instead consider the implication of their inclusion as part of our discussion.

The second survey sample we make use of is the Satellites Around Galactic Analogs (SAGA) survey \citep{GehaSAGA,2021Mao}.\ The SAGA survey aims to catalogue satellites as faint as Leo I $(M_r<-12.3)$ around approximately 100 nearby (D$\sim25-40.75$ Mpc) Milky Way analogs.\ These systems are selected primarily on their $K-$band luminosities $(-23<M_K<-24.6)$, however, there are a handful of secondary criteria, including halo mass and relative isolation (see \citealt[][]{2021Mao} for more details).\ Satellite candidates are spectroscopically confirmed to be associated with their putative hosts.\ The second stage release of the SAGA survey has catalogued 127 satellites around 36 hosts.\ The photometry for each of the SAGA satellites stems from extant photometric catalogues including SDSS DR14, DES DR1, and Legacy Survey DR6/DR7.\ We adopt their compiled photometric properties and discuss any potential limitations of these data in Section \ref{sec:summary}.\

The final set of satellites we use in this work comes from \citet[][hereafter N22]{2022Nashimoto} who search for satellites around 9 nearby ($15-20$ Mpc) Milky Way-like hosts using deep optical imaging from the Hyper Suprime-Cam on the Subaru Telescope.\ The Milky Way-like hosts are selected using infrared photometry from 2MASS as proxies for the stellar masses and imposing a halo mass range of $M_{halo} = (0.5 - 4) \times10^{12}M_{\odot}$ (see \citealt{2018Tanaka} for details) akin to the methodologies of the ELVES and SAGA surveys.\ A total of 93 satellites projected within their putative host's virial radius were determined to lie at similar distances via SBF distance estimates.\ 51 satellites were classified as 'secure' based on extant redshift measurements or based on their visual inspection to select candidates with smooth morphologies.\ The remaining 42 satellites were classified as 'possible' due to their smaller sizes and structure indicative of spiral arms or tidal features (see \citealt{2018Tanaka} for details).\ As with the ELVES catalogue, we consider the 51 satellites in this work as part of our primary analysis and discuss the implications of the broader sample afterward.\ This catalogue is considered to be complete down to $M_V\sim-10$ based on artificial dwarf injection and recovery tests \citep{2018Tanaka}, similar to those from \citet{ELVESI}.\ We note that for consistency with the other two samples we use the $K-$band magnitudes from \citet[][]{2017Kourkchi} in this work instead of the stellar and halo masses derived by \citetalias{2022Nashimoto}.\

There are a total of 75 Milky Way-like hosts with 516 confirmed satellites across all three survey samples.\ While there are certainly some differences between these three surveys (e.g.\ survey depths, candidate selection, confirmation methods, completeness limits, etc.), a fundamental motivator for these surveys is to characterize the satellite population of Milky Way analogs.\ We first consider these survey catalogues at face value for the analysis conducted in this work and then we discuss the implications of their differences in Section \ref{sec:summary}.

\subsection{Selecting UDGs}\label{subsec:udgs}
To select UDGs from the survey catalogues, we follow criteria from earlier works (i.e. \citealt[][]{2016vanderBurg,2017vanderBurg,roman2017,2019ManceraPina}).\ We impose a mean $g-$band effective surface brightness threshold of $\langle\mu_{g,\mathrm{eff}}\rangle \geq24 \,\mathrm{mag \, arcsec^{-2}}$ and a physical effective radius of $R_{\mathrm{eff}}\geq1.5\,\mathrm{kpc}$.\ We note that this surface brightness limit is brighter than the more commonly used central surface brightness limit of $\mu_{g,\mathrm{0}} \geq24 \,\mathrm{mag \, arcsec^{-2}}$ \citep[][]{2015vanDokkum}.\ Our criteria match those from some earlier works \citep[e.g.][]{2019ManceraPina}.\ However, they are only broadly consistent with those used by \citet[][]{2016vanderBurg,2017vanderBurg} who use circularized effective radii (i.e.\ $R_{\mathrm{eff,circ}} = R_{\mathrm{eff}}\sqrt{b/a}$) versus the more common use of the semi-major axis as the (non-circularized) effective radius.\ Using the circularized effective radius can have a broad range of effects on UDG abundance as shown by \citet[][]{2019ManceraPina} from no change to a decrease by up to a factor of $\sim2$.\ These changes would have a marginal effect on our results and we discuss similar effects from incompleteness in Section \ref{sec:summary}. 

The SAGA catalogue provides $r-$band apparent magnitudes, $g-r$ colours, and $\langle\mu_{r,\mathrm{eff}}\rangle$.\ From these values, we are able to obtain effective radii and calculate $\langle\mu_{g,\mathrm{eff}}\rangle$.\ We take similar steps to derive effective radii and $\langle\mu_{g,\mathrm{eff}}\rangle$ for the \citetalias{2022Nashimoto} catalogue where $i-$band apparent magnitudes, $g-i$ colours, and $\langle\mu_{i,\mathrm{eff}}\rangle$ are provided.\ For both of these survey catalogues, we calculate $R_{\mathrm{eff}}$ assuming the distance to the satellite's host.\ The ELVES catalogue does not explicitly compile $\langle\mu_{g,\mathrm{eff}}\rangle$ values.\ We calculate $\langle\mu_{g,\mathrm{eff}}\rangle$ following the same method as the SAGA survey (see Equation 2 in \citealt[][]{2021Mao}) using the $g-$band apparent magnitudes and the angular effective radius which we calculate using the provided physical effective radius and the distance to the satellite's host.\ We note that some satellites from ELVES only have $V-$band photometry provided.\ Of these systems, Sgr dSph would be the only one to satisfy our UDG criteria.\ We estimate its $\langle\mu_{g,\mathrm{eff}}\rangle$ by converting the $V-$band magnitude to $g-$band using the relations in \citet[][]{2021Carlsten} assuming $g-r=0.45$ and converting the listed $R_{\mathrm{eff}}$ to angular units assuming a distance of 18 kpc \citep{2012McConnachie}.\

\begin{figure}
    \centering
    \includegraphics[width=\columnwidth]{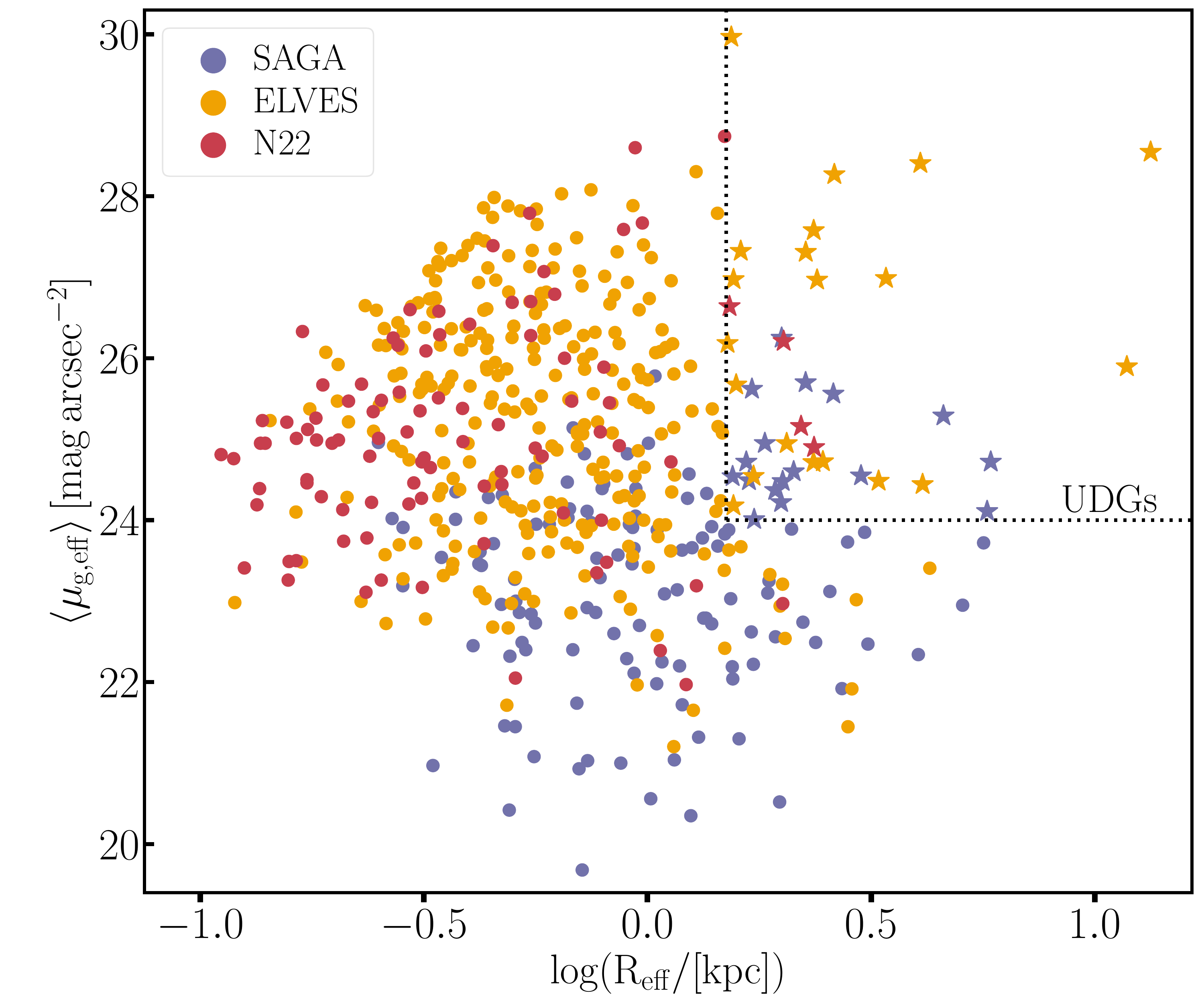}
    \caption{Mean $g-$band surface brightness within the effective radius as a function of physical effective radius for the three samples used in this work: SAGA (blue), ELVES (orange), and \citetalias{2022Nashimoto} (red).\ The parameter space where we select UDGs is defined by the dotted lines and satellites that satisfy these criteria are shown as stars while the rest are shown as circles.}
    \label{fig:rmu}
\end{figure}

In Figure \ref{fig:rmu}, we show $\langle\mu_{g,\mathrm{eff}}\rangle$ versus $R_{\mathrm{eff}}$ for all satellites.\ The SAGA, ELVES, and \citetalias{2022Nashimoto} satellites are coloured in blue, orange, and red, respectively.\ The dotted region in the upper-right of the figure shows the region in which UDGs reside and satellites satisfying the aforementioned thresholds are shown as stars.\ There are a total of 41 UDG satellites (17 SAGA, 20 ELVES, and 4 \citetalias{2022Nashimoto}) around 33 hosts (13 SAGA, 17 ELVES, and 3 \citetalias{2022Nashimoto}) across all three surveys.\ We note that while these 41 satellites do satisfy this \textit{mean} surface brightness criterion, half (20/41) do not satisfy the originally proposed central surface brightness criterion, $\mu_{g,0}\gtrsim24\,\mathrm{mag\,arcsec^{-2}}$ \citep{2015vanDokkum}.\ Furthermore, these UDG counts and the sample distributions in the $\langle\mu_{g,\mathrm{eff}}\rangle-R_{\mathrm{eff}}$ plane do not account for the differences between the surveys from which they are drawn.\ A key difference between them is the brighter SAGA absolute magnitude limit $(M_r <-12.3)$, which would exclude the majority of the faint satellites in the other samples.\ The relative abundance of UDG to non-UDG satellites is high for the SAGA sample $(17/127\sim13\%)$ compared to ELVES $(20/338\sim6\%)$ and \citetalias{2022Nashimoto} $(4/51\sim8\%)$.\ Accounting for this brighter absolute magnitude limit, the relative abundance of UDG to non-UDG satellites in ELVES  and \citetalias{2022Nashimoto} increases to $17/159\sim11\%$ and $4/28\sim14\%$, respectively.\ Indeed, the driver of this change is the significant decrease in the number of non-UDG satellites, which results in much more comparable values to SAGA.\ This decrease in non-UDG satellites would also manifest as a shift in the offset distributions in Figure \ref{fig:rmu} toward brighter and larger values (i.e.\ down and to the right).\ The ELVES median surface brightness and size shift from $\langle\mu_{g,\mathrm{eff}}\rangle=25.4\,\mathrm{mag\,arcsec^{-2}}$ and $R_{\mathrm{eff}}=0.58$ kpc to $\langle\mu_{g,\mathrm{eff}}\rangle=24.3\,\mathrm{mag\,arcsec^{-2}}$ and $R_{\mathrm{eff}}=0.85$ kpc. We see a similar shift in the \citetalias{2022Nashimoto} sample moving from $\langle\mu_{g,\mathrm{eff}}\rangle=24.9\,\mathrm{mag\,arcsec^{-2}}$ and $R_{\mathrm{eff}}=0.31$ kpc to $\langle\mu_{g,\mathrm{eff}}\rangle=24.4\,\mathrm{mag\,arcsec^{-2}}$ and $R_{\mathrm{eff}}=0.78$ kpc, although the sample size is relatively smaller.\ After the application of the absolute magnitude cut, both samples are much closer to the locus of the SAGA sample ($\langle\mu_{g,\mathrm{eff}}\rangle=23.5\,\mathrm{mag\,arcsec^{-2}}$, $R_{\mathrm{eff}}=1.01$ kpc).

\subsection{Estimating Host Masses}\label{subsec:hosts}
Previous studies of UDG abundances in groups and clusters estimated host halo masses using the velocity dispersions \citep[e.g.][]{roman2017,2018ManceraPina} or the total group $r-$band luminosity \citep[e.g.][]{2017vanderBurg}.\ We use an alternative method to estimate the halo masses of these Milky Way-like hosts and leverage the fact that all three surveys included in this work select their hosts primarily on their $K-$band luminosity.\ We use the Milky Way-like galaxies from the ARTEMIS suite of simulations from \cite{2020Font} to derive halo mass estimates for the hosts in our sample.\ The ARTEMIS suite contains 45 systems which have been demonstrated to match Milky Way-like stellar masses, sizes, and star formation rates \citep{2020Font,2021Font}.\ Furthermore, the distribution of ARTEMIS host $K-$band luminosities was shown to broadly agree with that of the SAGA survey \citep{2021Mao} and hosts from an earlier iteration of the ELVES survey \citep{2021Carlsten}.\ Given this consistency, we fit a linear relation between the $K-$band absolute magnitudes and the halo masses, $M_{200}$ for these ARTEMIS hosts.\ We note that we exclude one of these systems (G36) from this procedure as it is an outlier in this plane.\ This best-fit relation, $\mathrm{log}(M_{200}/M_{\odot}) = -0.19\cdot(M_K) + 7.45$, was used to estimate the halo masses of the Milky Way-like host galaxies in the observational samples considered here.\ Given the approximate nature of this method, we assume uncertainties of 0.5 dex for these halo mass estimates.

\begin{figure}
    \centering
    \includegraphics[width=\columnwidth]{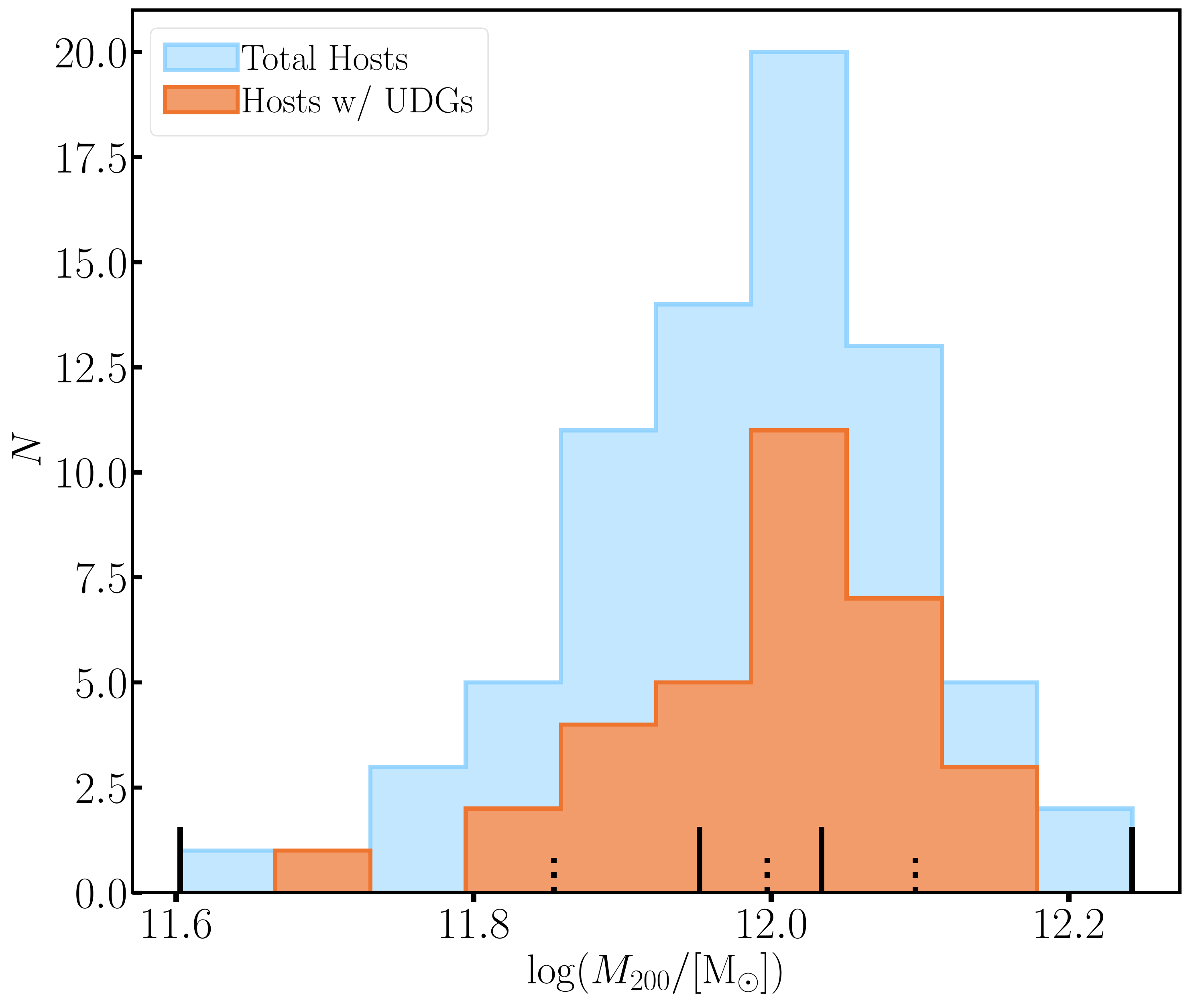}
    \caption{Distribution of host masses, $\mathrm{log}(M_{200})$.\ The total host sample is shown in blue, while those with UDGs in their satellite systems are shown in orange.\ The short solid and dotted lines at the bottom of the figure respectively show the edges and mean values of the host halo mass bins we use for our UDG abundance analysis in Figure \ref{fig:udgfreq}.}
    \label{fig:hosthist}
\end{figure}

In Figure \ref{fig:hosthist}, we show the distributions of the derived halo masses for all hosts (blue histogram) and for those with at least one UDG satellite (orange histogram).\ Both distributions peak around $\mathrm{log}(M_{200}/M_{\odot}) \sim12.05$, with a mean abundance of $\sim0.5$ UDG satellites.\ We show the halo mass bins that we will use in our UDG frequency analysis as the solid vertical lines at the bottom of the figure with the mean mass shown as the dotted vertical lines.\ We note the inclusion of the aforementioned outlier would shift these mean values down by $\sim0.05$ dex and would not affect the conclusions of this work.

\section{Results}\label{sec:results}
With the UDG abundances and host halo masses in hand, we now compare these results to a selection from the literature and extend the $N_{UDG}-M_{200}$ relation to Milky Way-mass systems.\ In Figure \ref{fig:hosthist}, we showed the distribution of host halo masses and three halo mass bins (edges and means, solid and dashed vertical lines, respectively) that were selected to contain a roughly equal number of hosts, hence their unequal widths.\ We determine the mean number of UDGs per host within each of the halo mass bins.\ To estimate the uncertainties on the number of UDGs within each halo mass bin, we draw 1000 samples of 25 hosts from a Poisson distribution with mean values between 0.1 and 1 in increments of 0.01.\ For each ensemble of distributions, we determine the probability of the measured mean values falling above and below our fiducial value (i.e.\ our calculated mean in Table \ref{tab:udg_abund}).\ Using these probabilities, we find the corresponding mean values that bracket $\sim1\sigma$ around our fiducial value.\ This approach closely matches the standard Poisson uncertainty but results in asymmetric uncertainties more suitable for the zero-biased nature of UDG abundances.\ We summarize these values in Table \ref{tab:udg_abund}

\begin{table*}
    \centering
    \caption{Summary of Host Properties and UDG Abundances}
    \begin{tabular}{cccccccc}
    \hline
    Bin & $N_{Hosts}$ & $N_{ELVES}$ & $N_{SAGA}$ & $N_{N22}$ & Mean $M_{200}$ & $M_{200}$ Range & $N_{UDG}$ \\
    \hline
    MW-low & 25 & 9 & 11 & 5 & 11.85 & [11.60, 11.95] & $0.32\substack{+0.13\\-0.07}$\\
    MW-mid & 25 & 7 & 15 & 3 & 12.00 & (11.95, 12.03] & $0.64\substack{+0.16\\-0.10}$\\
    MW-high & 25 & 14 & 10 & 1 & 12.10 & (12.03, 12.24] & $0.68\substack{+0.17\\-0.11}$\\
    \hline
    \end{tabular}

    \label{tab:udg_abund}
\end{table*}

\begin{figure*}
    \centering
    \includegraphics[width=\textwidth]{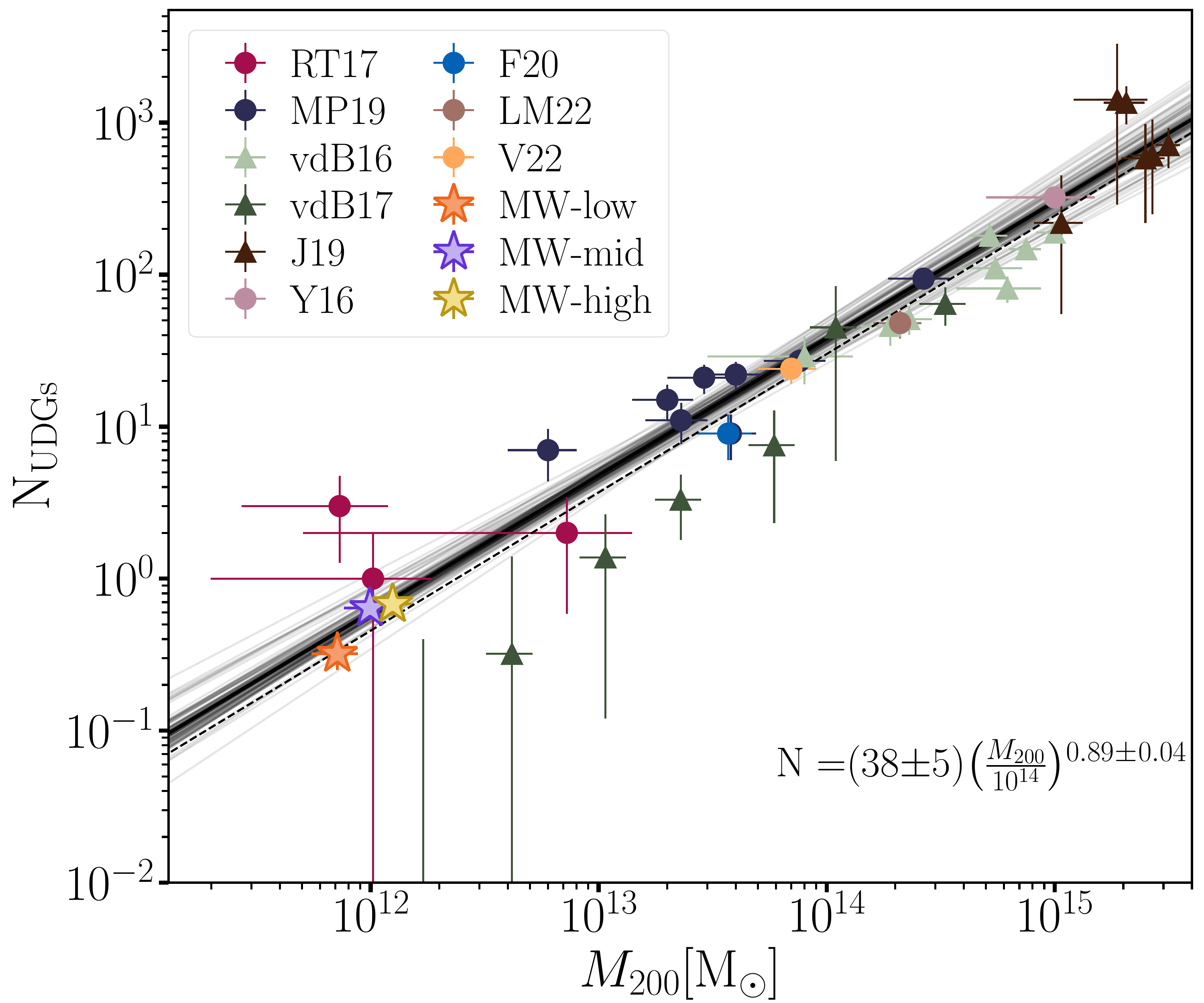}
    \caption{Number of UDGs as a function of halo mass.\ In addition to the Milky Way-mass hosts from this work (star symbols), we show the group and cluster samples included in this analysis as coloured triangles -- light green: \citet[][vdB16]{2016vanderBurg}, dark green: \citet[][vdB17]{2017vanderBurg}, and dark brown \citet[J19][]{2019Janssens} -- and circles -- red: \citet[][RT17]{roman2017}, dark blue: \citet[][MP19]{2019ManceraPina}, mauve: \citet[][Y16]{2016Yagi}, blue: \citet[][F22]{2020Forbes}, brown: \citet[][LM22]{2022LaMarca}, orange: \citet[][V22]{2022Venhola}.\ We note that the error bars, particularly on $N_{UDGs}$, may be hidden by the symbols.\  The solid black line shows our best-fit relation to all points excluding \citet[][]{2016vanderBurg,2017vanderBurg} and \citet{2019Janssens} samples along with the bootstrapped relations in grey, and the equation is shown at the top.\ We also show the best-fit relation when the \citet[][]{2016vanderBurg,2017vanderBurg} and \citet{2019Janssens} are included as the dashed black line.\ }
    \label{fig:udgfreq}
\end{figure*}

We show the number of UDGs as a function of host halo mass in Figure \ref{fig:udgfreq}.\ The three mass bins from this work are shown as the orange (MW--low), lavender (MW--mid), and yellow (MW--high) stars and the UDG abundances are the mean in each halo mass bin.\ The early work from \citet[][]{2016vanderBurg} for UDGs in galaxy clusters is shown in light green.\ We show the average group and cluster abundances from the \citet[][]{2017vanderBurg} extension of their previous work in dark green.\ We note that here we show the 0.1 dex errors on the mean halo masses adopted by \citet[][]{2017vanderBurg} when performing their fitting procedure.\ We also show the UDG abundances from the Coma cluster \citep[][]{2016Yagi}\footnote{We note that we use their catalogue and select the subset that satisfies our previously described criteria.}, Hickson Compact Groups \citep[][]{roman2017}, the KIWICS sample \citep{2018ManceraPina,2019ManceraPina}\footnote{Specifically, their values assuming non-circularized effective radii.}, the Frontier Field clusters \citep{2019Janssens}, the IC1459 group \citep{2020Forbes}, the Hydra I cluster \citep{2022LaMarca}, and the Fornax cluster \citep{2022Venhola}.\ 

We first consider the samples with relatively nearby hosts and non-circularized effective radii, excluding the \citet[][]{2016vanderBurg,2017vanderBurg} samples since these samples impose more restrictive selection criteria.\ To determine the best-fitting relation to these samples, we perform Orthogonal Distance Regression (ODR) and perform 1000 bootstraps resampling iterations, taking the mean and the standard deviation for the resulting slope and normalization constant from this procedure as our adopted values and associated uncertainties.\ Our best-fit relation is ${N_{UDG}}=38\pm5\cdot\big(\frac{M_{200}}{10^{14}}\big)^{0.89\pm0.04}$, which we show at the top of Figure \ref{fig:udgfreq} and is represented by the solid black line along with the bootstrap iterations shown as the light grey lines.\ This sub-linear relation is consistent with those from \citet[][]{roman2017} and \citet[][]{2019ManceraPina}.\

We repeat the fitting procedure for a second time now including group and cluster UDG abundances from \citet[][light and dark green circles]{2016vanderBurg,2017vanderBurg}.\ We also include the cluster UDG abundances from \citet[][grey circles]{2019Janssens} in this second iteration with the caveat that it is possible there may have been some evolution in these abundances since $z\sim0.3-0.5$.\ The resulting best-fit relation is ${N_{UDG}}=30\pm2\cdot\big(\frac{M_{200}}{10^{14}}\big)^{0.92\pm0.03}$, which is shown as the dashed black line in Figure \ref{fig:udgfreq}.\ The slopes from these two fitting procedures are consistent within their uncertainties.\ Similarly, this second best-fit relation is broadly consistent with earlier sub-linear $N_{UDG}-M_{200}$ relations and with the marginally super-linear relation from \citet[][]{2017vanderBurg}.\

\section{Discussion and Summary}\label{sec:summary}
We have explored the abundance of UDG satellites around Milky Way analogs using the SAGA, ELVES, and \citetalias{2022Nashimoto} samples.\ We select UDGs in a similar manner as done in previous studies for a total of 41 UDGs around 33 hosts out of the total 75 hosts.\ We group these hosts by their halo masses, taking the mean UDG abundance amongst them allowing us to extend the $N_{UDG}-M_{200}$ relation toward Milky Way halo masses.\

Our resulting best-fit relations in the $N_{UDG}-M_{200}$ plane have a sub-linear slope whether we exclude the groups and clusters from \citet[][]{2016vanderBurg,2017vanderBurg} (slope$=0.89\pm0.04$) or include them ($0.92\pm0.03$).\ The latter of these slopes is broadly consistent (within $\sim2\sigma$) with the super-linear slope ($1.11\pm0.07$) determined by \citet[][]{2017vanderBurg}.\ The sub-linear slope from our initial fitting procedure is consistent with the abundance of brighter dwarf galaxies in groups and clusters \citep[][]{2009TrenthamTully,2017vanderBurg}.\ Interestingly, our best-fit relation that does not include the \citet[][]{2016vanderBurg,2017vanderBurg} samples agrees, within uncertainties, with the abundance relation of brighter group and cluster members determined by \citet[][]{2017vanderBurg}.\ While this particular consistency may be a result of various systematic effects (e.g.\ satellite selection or fitting procedures), it implies that UDGs and brighter dwarfs likely follow similar abundance relations.\ 

As noted in earlier studies, a sub-linear $N_{UDG}-M_{200}$ relation has a couple of different interpretations and implications.\ A sub-linear relation excludes UDG formation mechanisms that require cluster environments since UDGs are less abundant per unit halo mass within them (also see the discussion in these earlier works).\ With the present data, it is difficult to statistically exclude specific mechanisms.\ However, it is clear that these data are well-fit with a single-power law over the full range (i.e.\ over 3 orders of magnitude) of halo mass.\ This suggests that UDG formation models that heavily rely on cluster environments (e.g.\ galaxy-galaxy harassment) are strongly challenged.\ Similarly, the agreement in the abundance relation between dwarfs and UDGs supports the lack of environment in determining UDG properties.\

We can also begin to determine the frequency of UDGs actively interacting with hosts or UDGs using signs of recent interactions.\ To accomplish this, we visually inspect the image of the 36 UDG satellites to search for signs of interactions.\ We find seven potentially interesting UDG satellites (ELVES: SGR, dw0240p3903, dw1105p0006, dw1120p1337, dw1123p1342, dw1908m6343; \citetalias{2022Nashimoto}: Obj.\ ID 115641) and most project near ($d_{\mathrm{proj}}<35$) kpc their hosts or a more massive companion\footnote{dw1123p1342 projects $\sim130$ kpc from its massive companion, however, it appears to fall along a faint stream connected to dw1120p1337.}.\ The tidal features and/or extended morphologies in these UDGs may be a result of recent and/or ongoing interactions with them.\ Further study of these systems may provide additional insight towards UDG formation via interactions \citep[][]{2018Bennet,carleton2019,tremmel2020, 2021Jones}.\ 

In their study of UDG satellites around isolated Milky Way-like hosts using the Auriga simulations, \citet[][]{liao2019} find a mean UDG abundance of $1.27 \pm 1.06$.\ This is well within the uncertainty of our abundances (see Table \ref{tab:udg_abund}) prior to accounting for the completeness corrections of our observed samples, which would increase our measured abundances.\ Their analysis was also able to distinguish satellites that formed as UDGs and those that formed through tidal interaction.\ They found that $\sim45\%$ of their UDGs are formed through a tidal interaction.\ While this is a higher fraction than in the observed samples considered here ($\sim20\%$), this difference is likely due to the ability of simulations to trace the entire evolutionary histories of these satellites as opposed to visual inspection of their imaging.\ The general agreement between these observed samples and simulations is interesting and warrants additional investigation.

Another recent study of note is the extensive search for UDGs and ``Ultra-Puffy Galaxies'' (UPGs) around Milky Way analogs conducted by \citet{2022Li} using imaging from the Hyper-Suprime Cam Strategic Survey Program \citep{2018Aihara}.\ After searching the projected virial radii of 689 Milky Way analogs, \citet{2022Li} find 412 UDGs around 258 hosts and a mean UDG abundance of $N_{UDG} = 0.44 \pm 0.05$ after applying background contamination and completeness corrections.\ While their surface brightness criterion is fainter than ours ($\langle\mu_{g,\mathrm{eff}}\rangle=25\,\mathrm{vs.\, 24 \, mag \, arcsec^{-2}}$), the calculated UDG abundances are still broadly consistent.\ Similarly, they find a marginally sub-linear slope of $0.96\pm0.04$ for the $N_{UDG}-M_{200}$ relation which is also consistent with our fiducial result within uncertainties.

A potential systematic that we have not accounted for is the completeness of the surveys used in our sample as well as the literature samples used in our analysis.\ Given the heterogeneous construction of these samples, it is difficult to account for all of their systematic differences.\ For example, the nearby Milky Way samples, particularly the ELVES and \citetalias{2022Nashimoto}, are relatively more complete in surface brightness when compared to the more distant group and cluster samples.\ This may account for additional UDG satellites that are included in these Milky Way samples.\ To test this, we impose an additional surface brightness threshold of $\langle\mu_{g,\mathrm{eff}}\rangle \leq26.5 \,\mathrm{mag \, arcsec^{-2}}$ for the UDG satellites from these samples.\ This limit is similar to the typical completeness of the group and cluster samples from the literature \citep[e.g.][]{2017vanderBurg,2019ManceraPina,2022Venhola}.\ We repeat our fitting procedure with these restricted Milky Way samples and those from the literature but excluding the \citet[][]{2016vanderBurg,2017vanderBurg} and \citet{2019Janssens} samples.\ We find a marginal increase in the best-fit slope $(0.94\pm0.05)$.\ Given that the completeness limits quoted in some of the literature samples are at the 50\% level, another test is to double their quoted $N_{UDG}$ values and perform our fits using their scaled $N_{UDG}$.\ The resulting best-fit slope $(0.97\pm0.07)$ does increase and while higher than our fiducial fit, it is now more consistent with a linear slope and the slightly super-linear slope of \citet[][]{2017vanderBurg}.\ These tests, however, assume that our Milky Way samples are themselves complete.

The surveys from which we draw our UDG satellites are not without their own incompleteness.\ There are an additional 148 satellite candidates within the ELVES and \citetalias{2022Nashimoto} samples that have tentative associations, of which 3 satisfy our UDG criteria.\ The SAGA sample has been suggested to be potentially missing low surface brightness satellites in their photometric catalogues \citep[][]{ELVESI,2022Font}, some of which may be UDGs.\ Combining this with the SAGA estimates of 24 faint satellites may have been missed during their spectroscopic follow-up \citep[][]{2021Mao}, there could be additional UDG satellites across their 36 hosts.\ In total, the handful of potential UDG satellites in these samples could increase our mean abundances and correspondingly make the $N_{UDG}$ relation shallower.\

\begin{figure}
    \centering
    \includegraphics[width=\columnwidth]{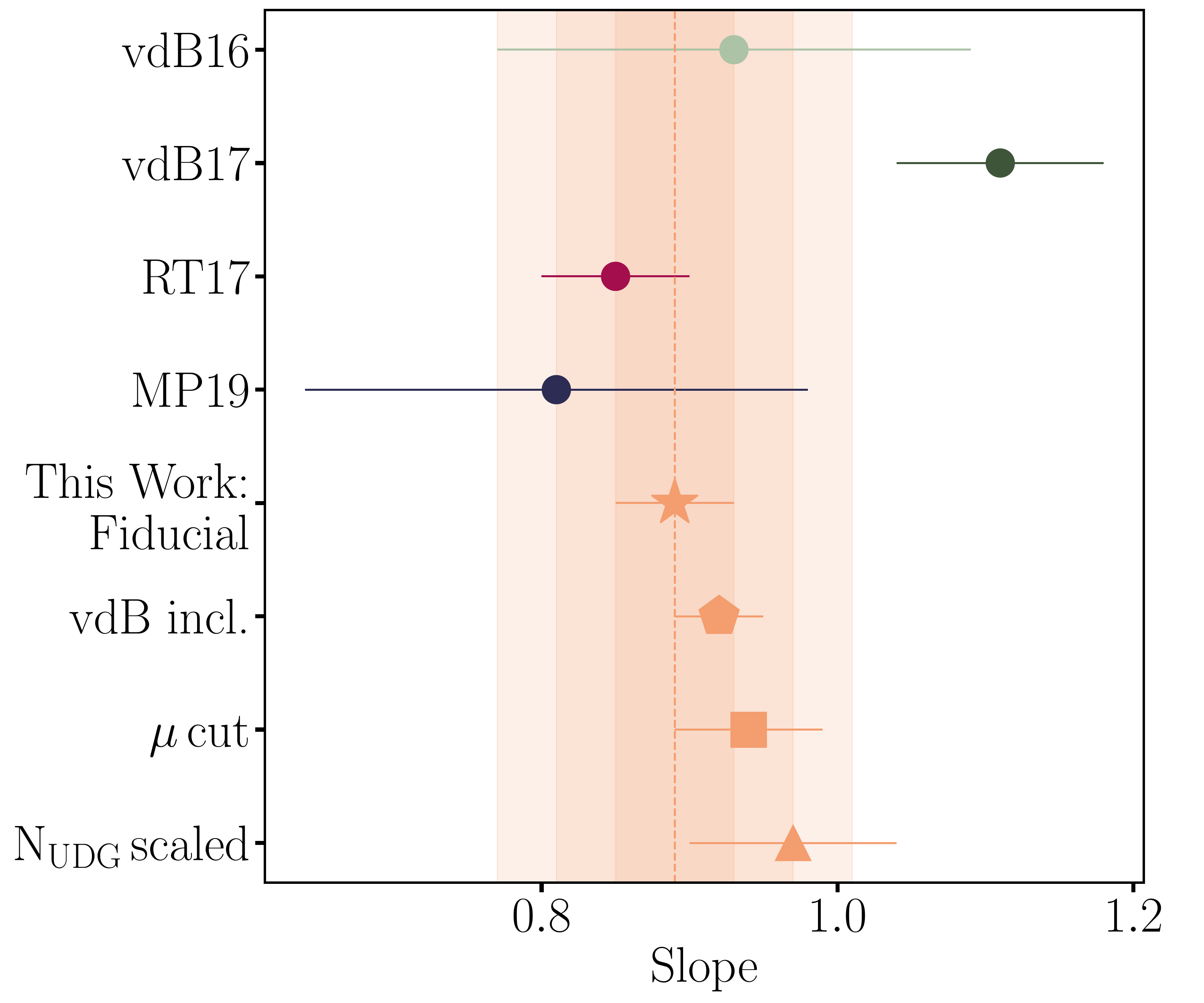}
    \caption{A comparison of the slopes from previous studies to those derived here.\ Literature slopes are shown as circles: \citet[][light green]{2016vanderBurg}, \citet[][dark green]{2017vanderBurg}, \citet[][red]{roman2017}, \citet[][dark blue]{2019ManceraPina}.\ Our best-fit slopes are shown in orange for different cases: fiducial (star), \citet[][]{2016vanderBurg,2017vanderBurg} and \citet[][]{2019Janssens} included (pentagon), surface brightness cut (square), and $N_{UDG}$ scaled (triangle).\ The shaded regions show the $1,\,2\,$and $3\sigma$ values based on our fiducial slope's uncertainty.}
    \label{fig:slopes}
\end{figure}

To facilitate a comparison of various best-fit slopes, we plot them together in Figure \ref{fig:slopes} with our slopes in orange and some of those from the literature coloured to match Figure \ref{fig:udgfreq}).\ The shaded regions show the $1$, $2$, and $3\sigma$ regions using our fiducial slope's uncertainty.\ We also refer the reader to \citet[][]{2018ManceraPina} where the authors uniformly derive UDG and host properties in addition to various cuts in UDG properties.\ Their resulting slopes are all consistent with those in Figure \ref{fig:slopes}.\ The results across samples, techniques, and environments are roughly consistent given their uncertainties.\ It is evident that progress, in terms of assessing what possible differences in slope (e.g.\ sub-linear or linear) might exist, will require significantly smaller uncertainties, and a fuller understanding of systems, which can only come from larger, homogeneous surveys of UDGs.

There is still much to learn about this intriguing subset of the LSB population and placing more stringent constraints on the abundance of UDGs around lower mass hosts is just one facet.\ With completed (i.e.\ ELVES) and ongoing (i.e.\ SAGA and \citetalias{2022Nashimoto}) surveys of Milky Way analogs in the Local Universe continuously providing growing samples, we can continue to characterize the UDG population around these hosts, particularly at lower host masses.\ Furthermore, these studies \citep[e.g.\ those used in this work and][]{2022Li} set the stage for future surveys such as LSST with the Rubin Observatory \citep{2019Ivezic}, which will expand the depths of these searches and will effectively help anchor the low mass end of the $N_{UDG}-M_{200}$ relation.\ Nevertheless, the near-linearity of the $N_{UDG}-M_{200}$ relation over more than 3 orders of magnitude in halo mass is a strong constraint on models of UDG formation and suggests that environmental factors are subdominant, although not necessarily negligible.\ Even a slope of 0.9, which one might consider near linear, would result in $\sim50\%$ fewer UDGs per unit halo mass in the most massive systems in these samples than in the least massive ones.

\section*{Acknowledgements}
We thank the anonymous referee for the insightful suggestions that improved the quality of this work.\  We also thank Shihong Liao and Steven Janssens for useful discussions.\ AK acknowledges financial support from the State Agency for Research of the Spanish Ministry of Science, Innovation and Universities through the "Center of Excellence Severo Ochoa" awarded to the Instituto de Astrof\'{i}sica de Andaluc\'{i}a (SEV-2017-0709) and through the grant POSTDOC$\_$21$\_$00845 financed from the budgetary program 54a Scientific Research and Innovation of the Economic Transformation, Industry, Knowledge and Universities Council of the Regional Government of Andalusia.\
This research made use of data from the SAGA Survey (sagasurvey.org). The SAGA Survey is a spectroscopic survey with data obtained from the Anglo-Australian Telescope, the MMT Observatory, and the Hale Telescope at Palomar Observatory. The SAGA Survey made use of public imaging data from the Sloan Digital Sky Survey (SDSS), the DESI Legacy Imaging Surveys, and the Dark Energy Survey, and also public redshift catalogs from SDSS, GAMA, WiggleZ, 2dF, OzDES, 6dF, 2dFLenS, and LCRS. The SAGA Survey was supported by NSF collaborative grants AST-1517148 and AST-1517422 and by Heising–Simons Foundation grant 2019-1402.

\section*{Data Availability}
All data used in this work are freely available from their respective original publications.



\bibliographystyle{mnras}
\bibliography{references} 

\bsp	
\label{lastpage}
\end{document}